\documentclass{aastex701}

\usepackage{algorithm}
\usepackage{algpseudocode}

\newcommand{\naoc}{National Astronomical Observatories, Chinese Academy of Sciences, Beijing 100101, China}
\newcommand{\ucas}{University of Chinese Academy of Sciences, Beijing 100049, China}
\newcommand{\sdju}{School of Arts and Sciences, Shanghai Dianji University, Shanghai 200240, China}
\newcommand{\qlnu}{College of Physics and Electronic Engineering, Qilu Normal University, Jinan 250300, China}
\newcommand{\nscc}{National Supercomputing Center in Jinan, Jinan 250103, China}
\newcommand{\tsinghuaastro}{New Cornerstone Science Laboratory, Department of Astronomy, Tsinghua University, Beijing 100084, China}
\newcommand{\sharedauthoremail}{\email{dingjialang25@mails.ucas.ac.cn}} 

\shorttitle{QUEST software correlator}
\shortauthors{Ding et al.}

\begin{document}

\title{A Python/CuPy Software Correlator for QUEST: Real-Time Performance and Initial Imaging}

\author[orcid=0009-0003-9837-1802]{Jialang Ding}
\affiliation{\naoc}
\affiliation{\ucas}
\email{dingjialang25@mails.ucas.ac.cn}

\author[orcid=0009-0006-0364-4161]{Guanhong Lin}
\affiliation{\naoc}
\affiliation{\ucas}
\sharedauthoremail

\author{Dejia Zhou}
\affiliation{\naoc}
\affiliation{\ucas}
\sharedauthoremail

\author[orcid=0000-0002-2940-4821]{Jianli Zhang}
\affiliation{\naoc}
\sharedauthoremail

\author{Ran Duan}
\affiliation{\naoc}
\sharedauthoremail

\author{Fei Liu}
\affiliation{\naoc}
\sharedauthoremail

\author{Xiaoyun Ma}
\affiliation{\naoc}
\sharedauthoremail

\author{Jie Zhang}
\affiliation{\sdju}
\affiliation{\qlnu}
\sharedauthoremail

\author{Meng Liu}
\affiliation{\qlnu}
\sharedauthoremail

\author{Chenchen Miao}
\affiliation{\qlnu}
\sharedauthoremail

\author{Yuan Liang}
\affiliation{\naoc}
\affiliation{\ucas}
\sharedauthoremail

\author{Liaoyuan Liu}
\affiliation{\naoc}
\affiliation{\ucas}
\sharedauthoremail

\author{Yingrou Zhan}
\affiliation{\naoc}
\affiliation{\ucas}
\sharedauthoremail

\author{Yuting Chu}
\affiliation{\naoc}
\affiliation{\ucas}
\sharedauthoremail

\author{Jing Qiao}
\affiliation{\naoc}
\affiliation{\ucas}
\sharedauthoremail

\author{Wei Wang}
\affiliation{\naoc}
\sharedauthoremail

\author{Zerui Wang}
\affiliation{\qlnu}
\sharedauthoremail

\author{Menquan Liu}
\affiliation{\sdju}
\affiliation{\qlnu}
\sharedauthoremail

\author{Meng Guo}
\affiliation{\nscc}
\sharedauthoremail

\author{Di Li}
\affiliation{\tsinghuaastro}
\affiliation{\naoc}
\sharedauthoremail

\author{Pei Wang}
\affiliation{\naoc}
\sharedauthoremail

\author{Xuanyu Wang}
\affiliation{\naoc}
\affiliation{\ucas}
\sharedauthoremail

\author{Xiaohui Yan}
\affiliation{\naoc}
\affiliation{\ucas}
\sharedauthoremail


\begin{abstract}
We present a Python/CuPy FX software correlator for small radio interferometer arrays and evaluate it on QUEST (Qilu University Explorer Survey Telescope). The system combines multi-threaded data ingest, pinned-memory host--device transfers, GPU-accelerated correlation, Polyphase Filter Bank channelization, MAD-based RFI flagging, and delay/phase calibration in a single workflow aimed at array commissioning. On a single NVIDIA RTX~4090D GPU, the implementation reaches a peak throughput of 1.51\,GB\,s$^{-1}$, which is sufficient for real-time operation in the four-antenna mode tested here. After calibration, the visibility phase across a clean 1.32--1.38\,GHz band is flattened to a residual scatter of a few degrees. Using the calibrated visibilities, we form a four-antenna synthesis image of Cassiopeia~A; the CLEANed image recovers a compact source at the phase center and reduces image-domain background fluctuations from order $10^{-1}$ to a few $10^{-2}$\,Jy\,beam$^{-1}$. These results indicate that the software is suitable for small-array commissioning and initial synthesis imaging on QUEST. A GNSS-based beam measurement is included as a supporting commissioning check.
\end{abstract}


\section{Introduction}
\label{sec:intro}

Modern radio interferometers rely on correlators that must sustain high data rates while preserving the phase information needed for calibration and imaging \citep{thompson2017interferometry}. For very large arrays, correlator development is typically driven by absolute throughput, power efficiency, and large-$N$ scalability, leading to heavily optimized FPGA, ASIC, and low-level GPU implementations \citep{clark2013xgpu,galaxies11010013}. Small arrays and prototype instruments often face a different set of priorities. In that regime, it is frequently more important to deploy quickly, inspect intermediate products easily, and modify the software during commissioning than to maximize raw performance at all costs.

This paper addresses that small-array regime. We present a Python/CuPy correlator and calibration workflow aimed at arrays of roughly 4--32 antennas, where software maintainability and commissioning utility are central design requirements. Rather than framing the system as a general replacement for large-array production correlators, we focus on a narrower question: can a high-level GPU-accelerated software correlator deliver calibrated visibilities that are good enough for real-time operation and initial synthesis imaging on a small interferometer?

For this paper, the main contributions are:
\begin{itemize}
    \item We present a GPU-accelerated FX correlator implemented in Python/CuPy, with multi-threaded I/O, pinned-memory transfers, and configurable PFB channelization.
    \item We characterize the practical operating point of the software on the QUEST platform and show that a single RTX~4090D can sustain real-time processing for the four-antenna configuration tested here.
    \item We implement a compact post-processing chain consisting of MAD-based RFI flagging together with coarse delay, fine delay, and residual phase correction, and we verify the resulting visibilities through synthesis imaging of Cassiopeia~A.
\end{itemize}

\begin{figure}[ht!]
\includegraphics[width=0.4\textwidth]{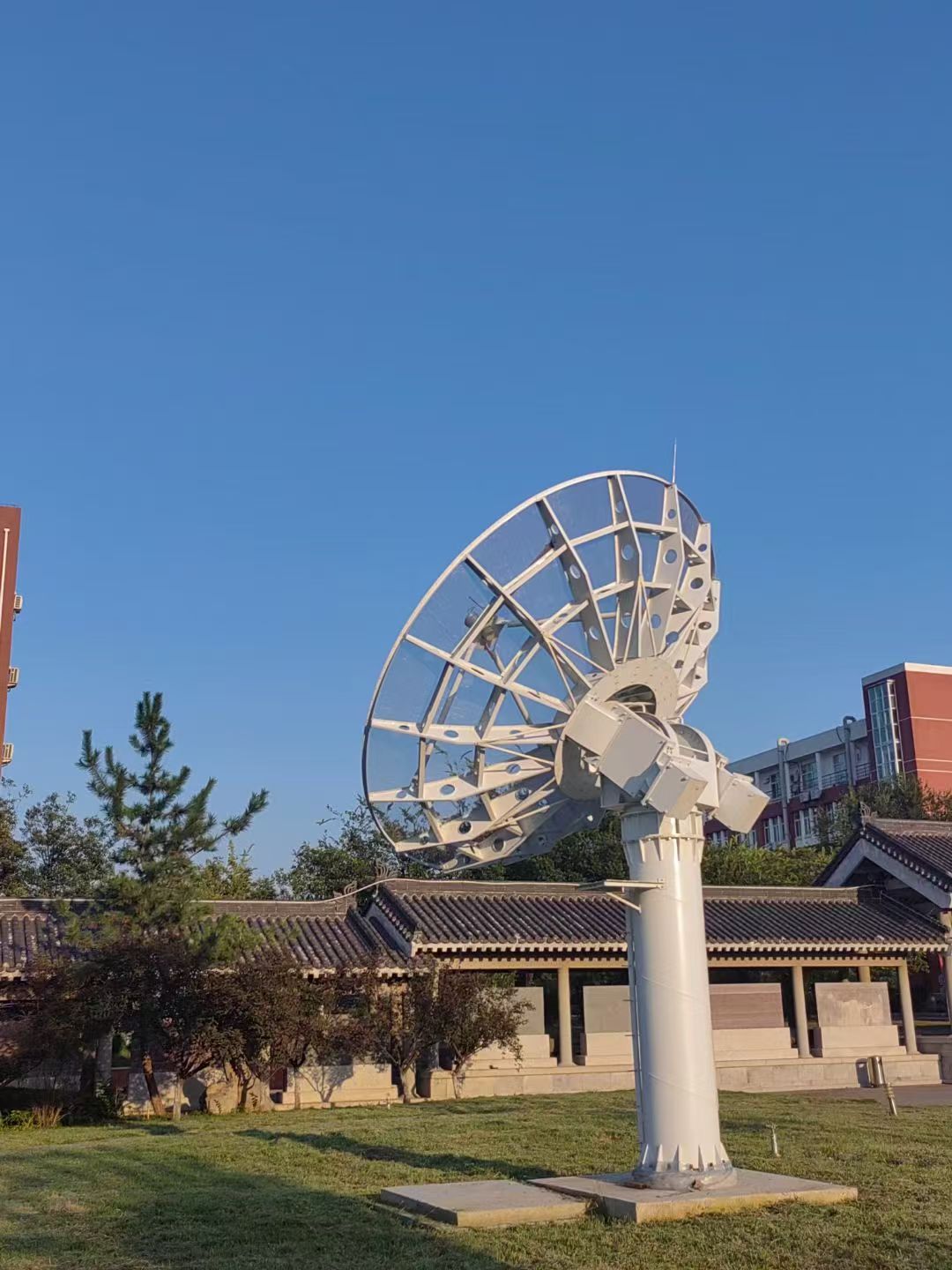}
\centering
\caption{The QUEST 4.5\,m diameter radio telescope.
\label{fig:QL}}
\end{figure}

The QUEST array serves as the testbed for this study (Figure~\ref{fig:QL}). Throughout the paper, the correlator itself and the Cassiopeia~A imaging demonstration form the main narrative. The RFI analysis and the GNSS beam experiment are included as supporting commissioning diagnostics rather than as standalone claims of equal weight. The remainder of the paper is organized as follows. Section~\ref{sec:instrument} summarizes the QUEST hardware. Section~\ref{sec:acquisition} describes the acquisition and recorder chain. Section~\ref{sec:correlator} presents the correlator architecture. Sections~\ref{sec:rfi}--\ref{sec:calibration} describe the supporting RFI analysis, performance characterization, and calibration workflow. Section~\ref{sec:imaging} presents the imaging validation, with Cassiopeia~A as the primary demonstration. Section~\ref{sec:conclusion} summarizes the main results and limitations.

\section{The QUEST Array}
\label{sec:instrument}

QUEST (Qilu University Explorer Survey Telescope) is a decimeter-wave radio interferometer located in Zhangqiu District, Jinan, Shandong Province. The full array currently comprises twenty antennas distributed across three sites: six antennas at the north gate, ten antennas at the east gate, and four antennas located at a supercomputing park approximately 20\,km away. For the present paper, QUEST is used as a commissioning testbed for the software correlator, and a four-antenna subarray is used for the end-to-end imaging demonstration in Section~\ref{sec:imaging}.

The antennas adopt a prime-focus revolving paraboloid design. Their main parameters are listed in Table~\ref{tab:antenna_params}. In the context of this paper, two quantities are especially relevant: the $\sim 3.5^\circ$ half-power beamwidth near 1.3\,GHz, which sets the scale for the GNSS beam-mapping experiment, and the dual-linear-polarization front end, which defines the correlator input streams.

\begin{deluxetable}{ll}[h!]
\tablecaption{Antenna Parameters \label{tab:antenna_params}}
\tablehead{\colhead{Parameter} & \colhead{Value}}
\startdata
D & 4.5m \\
F/D & 0.38 \\
HPBW & 3.5$^{\circ}$ @ 1.3GHz \\
BW & 0.8-1.7GHz \\
Polarization & Double linear polarization \\
Gain & 40dB \\
Pointing accuracy & 0.1$^{\circ}$ \\
Max tracking speed & 1$^{\circ}$/s \\
\enddata
\end{deluxetable}

\begin{figure*}[ht!]
\includegraphics[width=0.8\textwidth]{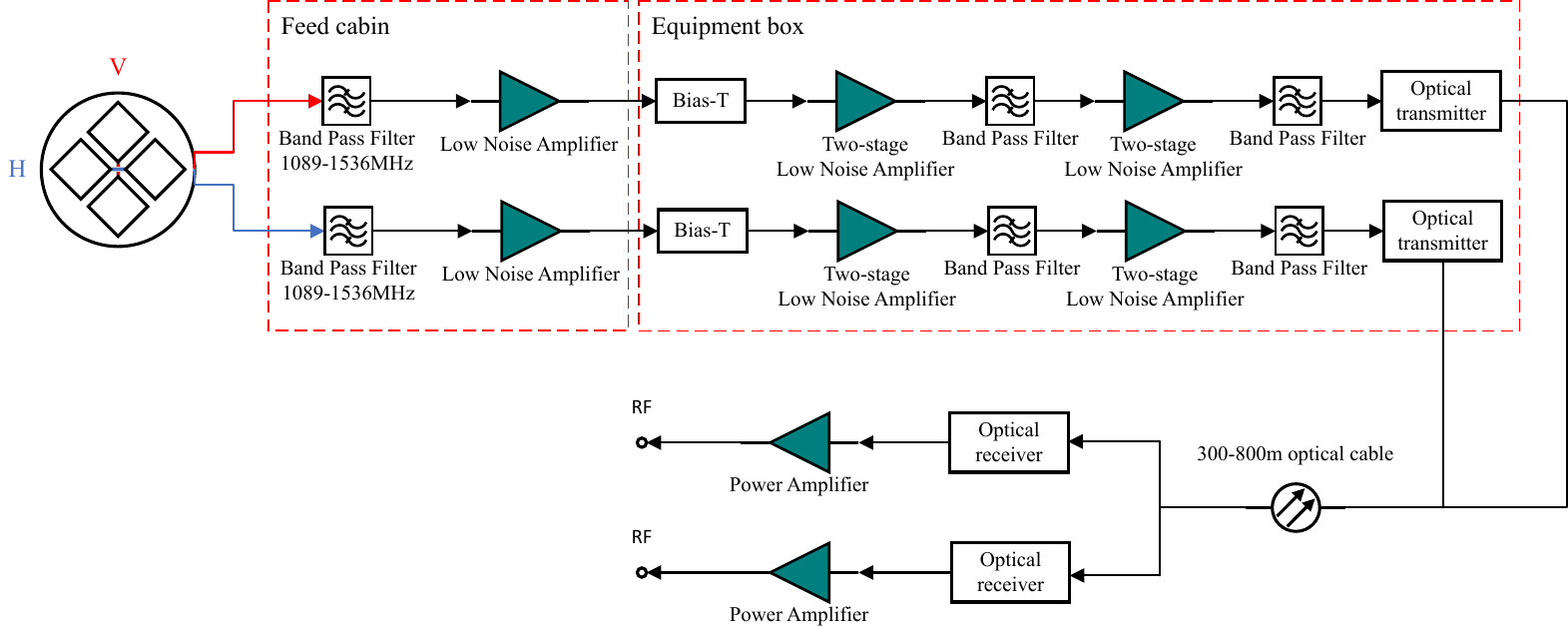}
\centering
\caption{RF front-end diagram.
\label{fig:rf_link}}
\end{figure*}

The RF front end for the sixteen on-campus antennas is shown in Figure~\ref{fig:rf_link}. A low-loss band-pass filter and the first-stage low-noise amplifier (LNA) are placed in the feed cabin. A DC Bias-Tee, a second amplifier stage, an additional band-pass filter, and an optical transmitter are integrated in an equipment box mounted beneath the antenna. Signals are then transported to the central equipment room via RF-over-Fiber (RFoF), a scheme widely used in radio instrumentation because it combines low transport loss with good immunity to electromagnetic interference over long baselines \citep{Virone2022,Mena2013}. Each on-campus antenna is connected to the central room through a dedicated 12-core optical fiber, and the unequal fiber lengths must later be accounted for in delay calibration. The equipment box and RFoF chain together provide more than 40\,dB of gain, while the downstream band-pass filter selects the 1050--1450\,MHz range.

\section{Data Acquisition System}
\label{sec:acquisition}

\begin{figure}[ht!]
\includegraphics[width=0.8\textwidth]{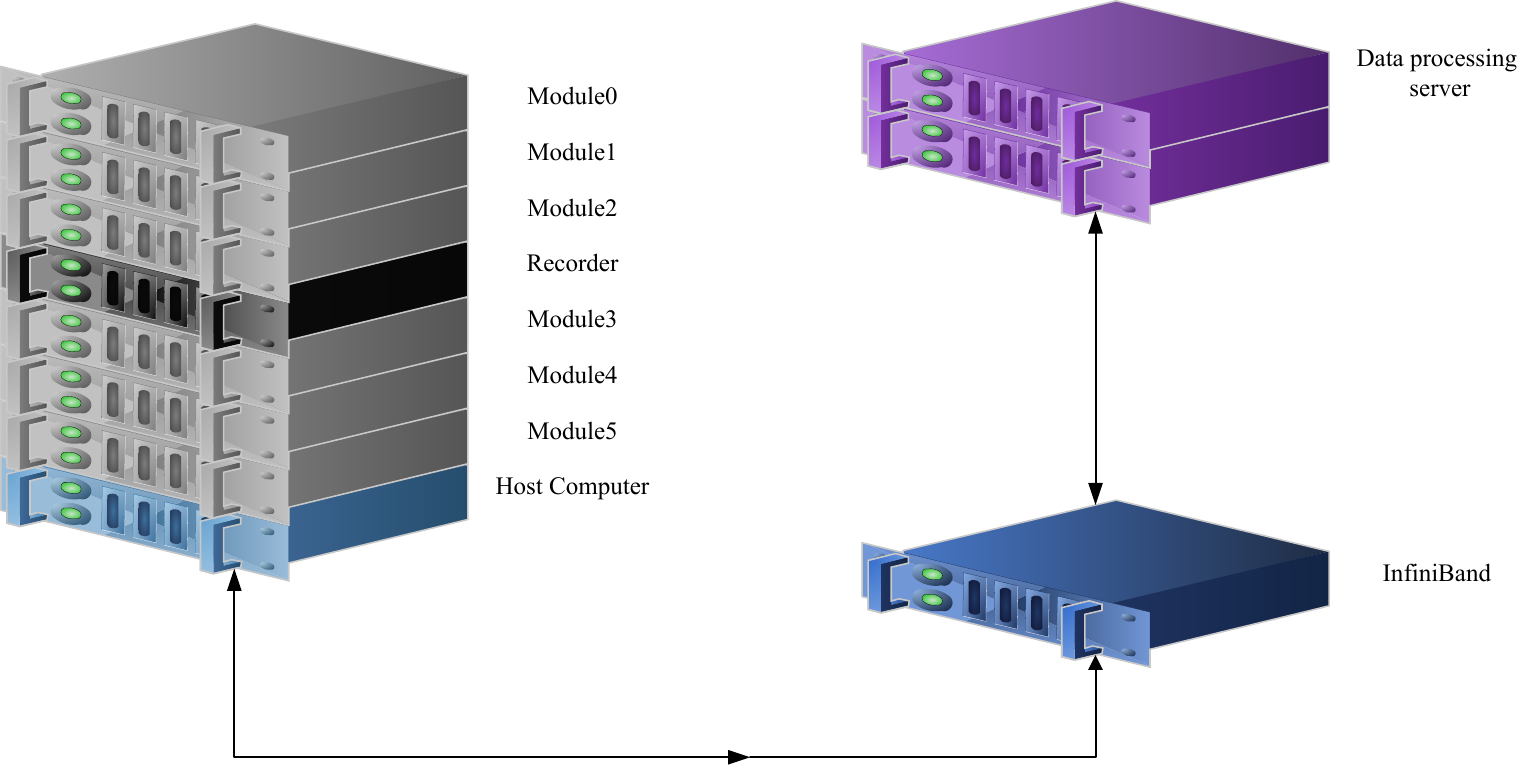}
\centering
\caption{Terminal network diagram.
\label{fig:network}}
\end{figure}

The backend is organized into an acquisition terminal and a processing terminal connected through high-speed InfiniBand (IB) networking \citep{clark2013xgpu}. The acquisition side comprises six identical RFSoC-based modules, a central recorder, and a host computer, while the processing side consists of two heterogeneous servers (Figure~\ref{fig:network}). This separation allows the acquisition hardware to remain focused on deterministic data capture while the software correlator runs on commodity compute servers.

\subsection{Acquisition Modules and Sampling}
Each acquisition module is built around a Zynq UltraScale+ RFSoC and performs direct RF sampling at 4000\,MHz. In the validated observing mode used throughout this paper, the firmware applies an $8\times$ decimation, yielding a 500\,MHz effective bandwidth centered at 1300\,MHz. At present, higher decimation factors for narrowband operation are not enabled in the production firmware because they still require additional verification against aliasing and mixing artifacts. Accordingly, the 500\,MHz mode is treated here as the reference operating mode.

\subsection{Data Transmission and the Central Recorder}
The six acquisition modules deliver their output to the central recorder over a 40\,Gbps Aurora link. The recorder, implemented on an FPGA platform, serves as the merger point for all module streams and can also perform several online logic operations, including secondary decimation without additional filtering, standard FFT, and integrated FFT modes. In practice, QUEST uses two complementary modes: high-rate baseband recording for flexible offline processing, and reduced data-rate spectral monitoring for rapid diagnostics.

\subsection{Quantization Logic and Bit-depth}
Quantization and bit-width conversion are handled in the recorder FPGA. To balance dynamic range against storage cost, the system currently supports two output modes:
\begin{itemize}
    \item \textbf{16-bit padded mode (default):} the native 14-bit ADC output is padded to 16 bits without truncation, preserving the full sampled dynamic range during transport and correlation.
    \item \textbf{8-bit truncated mode:} the output is truncated to 8 bits for data-volume-limited observations in which reduced precision is acceptable.
\end{itemize}

The main acquisition-system parameters are summarized in Table~\ref{tab:acq_params}. For the experiments reported in this paper, the data path is synchronized by a 1PPS plus 100\,MHz reference-clock distribution and forwarded to the processing server through a 100G QSFP28 link.

\begin{deluxetable}{ll}[h!]
\tablecaption{Acquisition System Specifications \label{tab:acq_params}}
\tablehead{\colhead{Parameter} & \colhead{Value}}
\startdata
Acquisition Board & Zynq UltraScale+ RFSoC \\
Sampling Rate & 4000 MHz (8x decimation) \\
Effective Bandwidth & 500 MHz \\
Inter-module Link & 40 Gbps Aurora \\
Recorder Functions & Aggregation, Decimation, FFT \\
Data Precision & 16-bit (padded) / 8-bit (truncated) \\
Synchronization & 1PPS + 100MHz Clock \\
Network Transfer & 100G QSFP28 (to IB) \\
\enddata
\end{deluxetable}

\section{The FX Correlator} \label{sec:correlator}

GPU-based correlators are now standard tools in radio astronomy, but their implementations span a wide range of design goals \citep{Barsdell2010,clark2013xgpu}. The correlator presented here is optimized for a different niche than large-array production systems. Its primary objective is to provide a code base that can be modified quickly during instrument commissioning while still sustaining real-time operation for a small-$N$ array. For that reason, the implementation is built in Python with CuPy acceleration \citep{Okuta2017}, rather than in a fully hand-optimized low-level CUDA code base.

\subsection{Mathematical Framework}
The FX correlator measures the cross-correlation between the electric-field voltages $V_i(t)$ and $V_j(t)$ received by antenna pair $(i,j)$. In the frequency domain the operation can be summarized as follows:
\begin{enumerate}
    \item The time-domain voltage stream $V(t)$ is transformed into $V(\nu)$ by the F-engine.
    \item The geometric delay $\tau_{g,ij}$ is compensated by applying a linear phase rotation $e^{j2\pi\nu\tau_{g,ij}}$.
    \item The cross-power spectrum is formed as
    \begin{equation}
    S_{ij}(\nu) = \left\langle V_i(\nu)\,V_j^*(\nu)\,e^{j2\pi\nu\tau_{g,ij}} \right\rangle ,
    \end{equation}
    where $V_j^*(\nu)$ is the complex conjugate of the signal from antenna $j$ and $\langle \cdot \rangle$ denotes temporal integration.
\end{enumerate}

\begin{figure*}[ht!]
\centering
\includegraphics[width=0.8\textwidth]{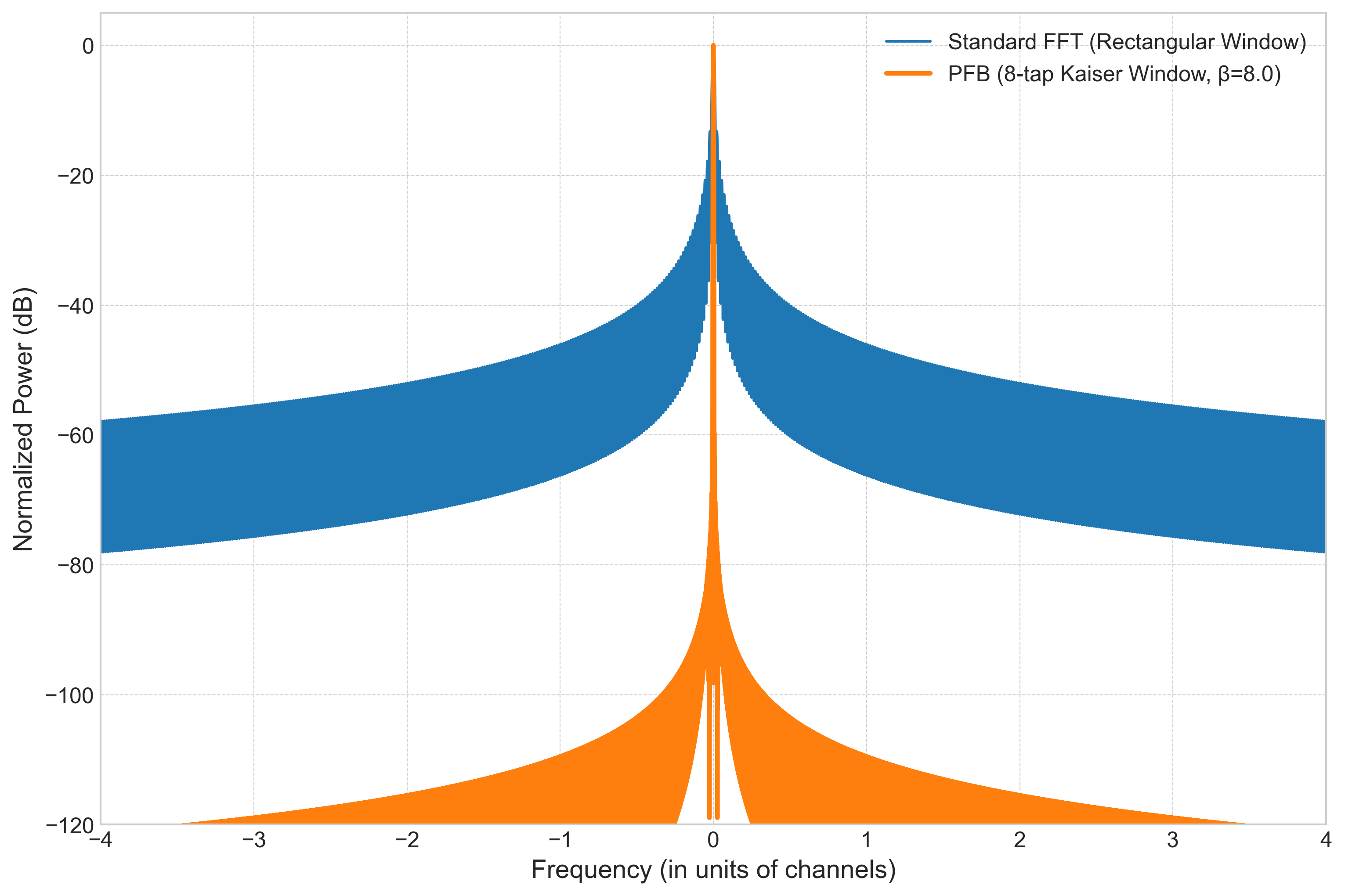}
\caption{Comparison of sidelobe suppression between a standard FFT (rectangular window) and an 8-tap PFB with a Kaiser window ($\beta=8.0$). The PFB reduces spectral leakage and is therefore preferred for the RFI-rich observing environment discussed in this paper.
\label{fig:pfb_fft}}
\end{figure*}

\begin{figure*}[h!]
\includegraphics[width=0.8\textwidth]{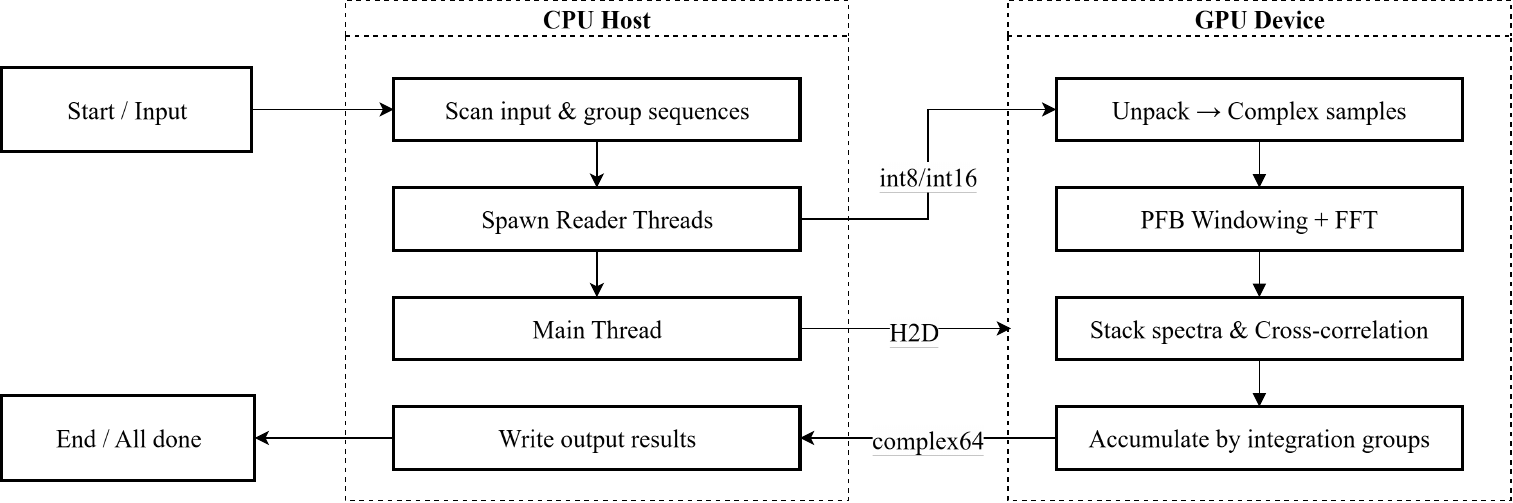}
\centering
\caption{Software framework flowchart of the correlator. CPU tasks are dominated by I/O and scheduling, while the GPU handles unpacking, channelization, cross-multiplication, and integration.
\label{fig:flowchart}}
\end{figure*}

\subsection{Pipeline Architecture and Implementation}
The software is organized to overlap disk/network I/O with GPU computation. Figure~\ref{fig:flowchart} summarizes the separation between host-side tasks and device-side tasks.

\textbf{Asynchronous I/O:} Reader threads follow a producer--consumer pattern and stream packets into pinned (page-locked) host memory. This choice enables asynchronous host-to-device transfers through DMA and reduces the scheduling overhead that would otherwise limit throughput on the CPU side.

\textbf{F-engine (PFB and FFT):} After transfer to the GPU, the raw baseband samples are unpacked and converted into complex voltage streams. Channelization is carried out with an FFT preceded by a PFB stage. In the RFI-rich QUEST environment, the improved sidelobe suppression of the PFB relative to a plain rectangular-window FFT is operationally important because it limits spectral leakage from strong narrow-band contaminants (Figure~\ref{fig:pfb_fft}).

\textbf{X-engine and integration:} The X-engine computes $V_iV_j^*$ for all baselines, and the resulting visibilities are accumulated in GPU memory over a user-defined integration interval before being copied back to the host and written to disk.

The overall program logic is summarized in Algorithm~\ref{alg:correlator}.

\begin{algorithm}[H]
\caption{Asynchronous FX Correlator Pipeline}\label{alg:correlator}
\begin{algorithmic}[1]
\State \textbf{Initialize} Reader threads and allocated Pinned Memory buffers
\While{Observation is active}
    \State \Comment{CPU Thread: Data Acquisition}
    \For{each antenna $i$}
        \State Read raw packets into \textit{PinnedBuffer\_i}
        \State Push buffer pointer to \textit{DataQueue} (up to \textit{queue-max})
    \EndFor

    \State \Comment{GPU Device: Processing Loop}
    \For{each \textit{batch-packets}}
        \State Pop buffers from \textit{DataQueue}
        \State \textbf{Async H2D:} Transfer data to GPU memory via \textit{Stream}
        \State \textbf{F-Engine:} Apply PFB weighting $\to$ FFT
        \State \textbf{X-Engine:} Compute $V_i \cdot V_j^*$ for all baselines
        \State \textbf{Integration:} Accumulate results in GPU global memory
    \EndFor
    \If{Integration period reached}
        \State \textbf{Async D2H:} Copy visibilities to host and save as FITS
    \EndIf
\EndWhile
\end{algorithmic}
\end{algorithm}

\begin{figure*}[h!]
\centering
\includegraphics[width=\textwidth]{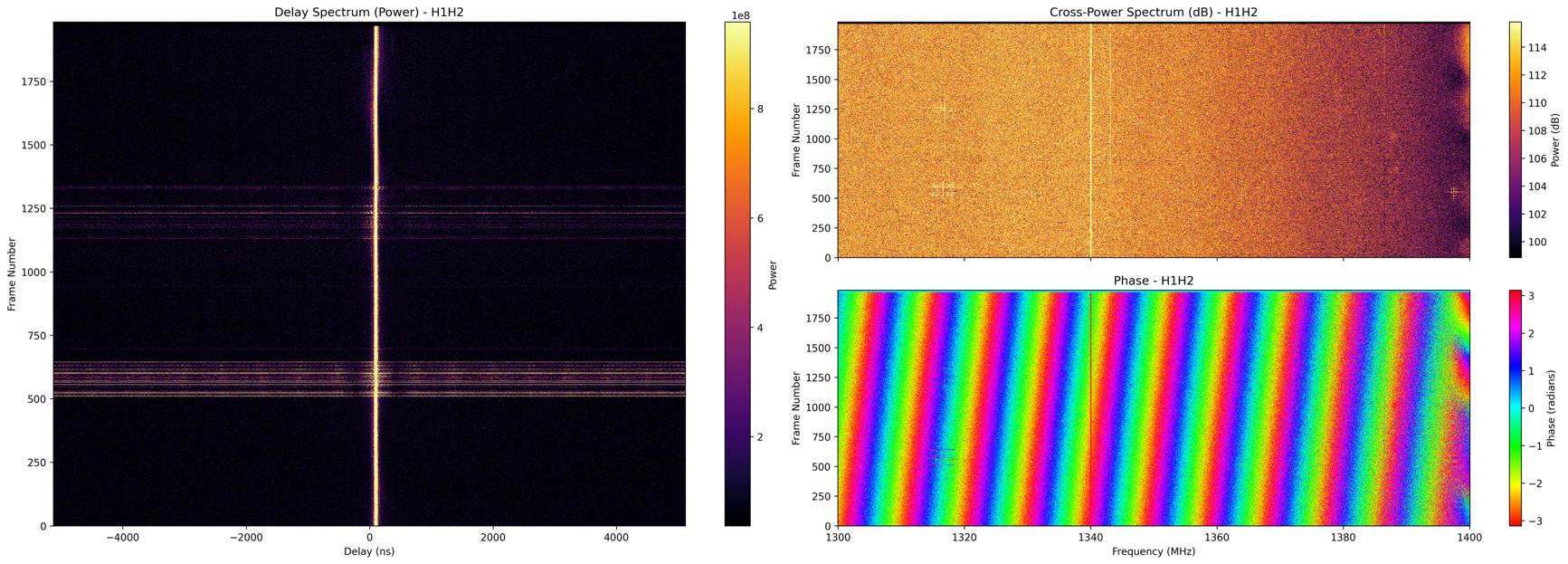}
\caption{Example correlator output from continuous tracking of Cas~A. The bandwidth is 1300--1400\,MHz and the integration time is about 0.27\,s per frame. The delay-spectrum peak remains stable over the short time interval shown, while the right-hand panels display the corresponding cross-power amplitude and phase.
\label{fig:fx_output_casa}}
\end{figure*}

\subsection{Flexibility and Scalability}
The framework exposes FFT length, number of PFB taps, integration interval, and queue depth as runtime parameters. This makes it useful both as an operational correlator and as a commissioning tool. In the present paper we focus on the validated small-array regime, where the software provides stable output products suitable for the RFI analysis, calibration, and imaging demonstrations discussed below.

\section{RFI Characterization} \label{sec:rfi}

\begin{figure*}[ht!]
\includegraphics[width=0.65\textwidth]{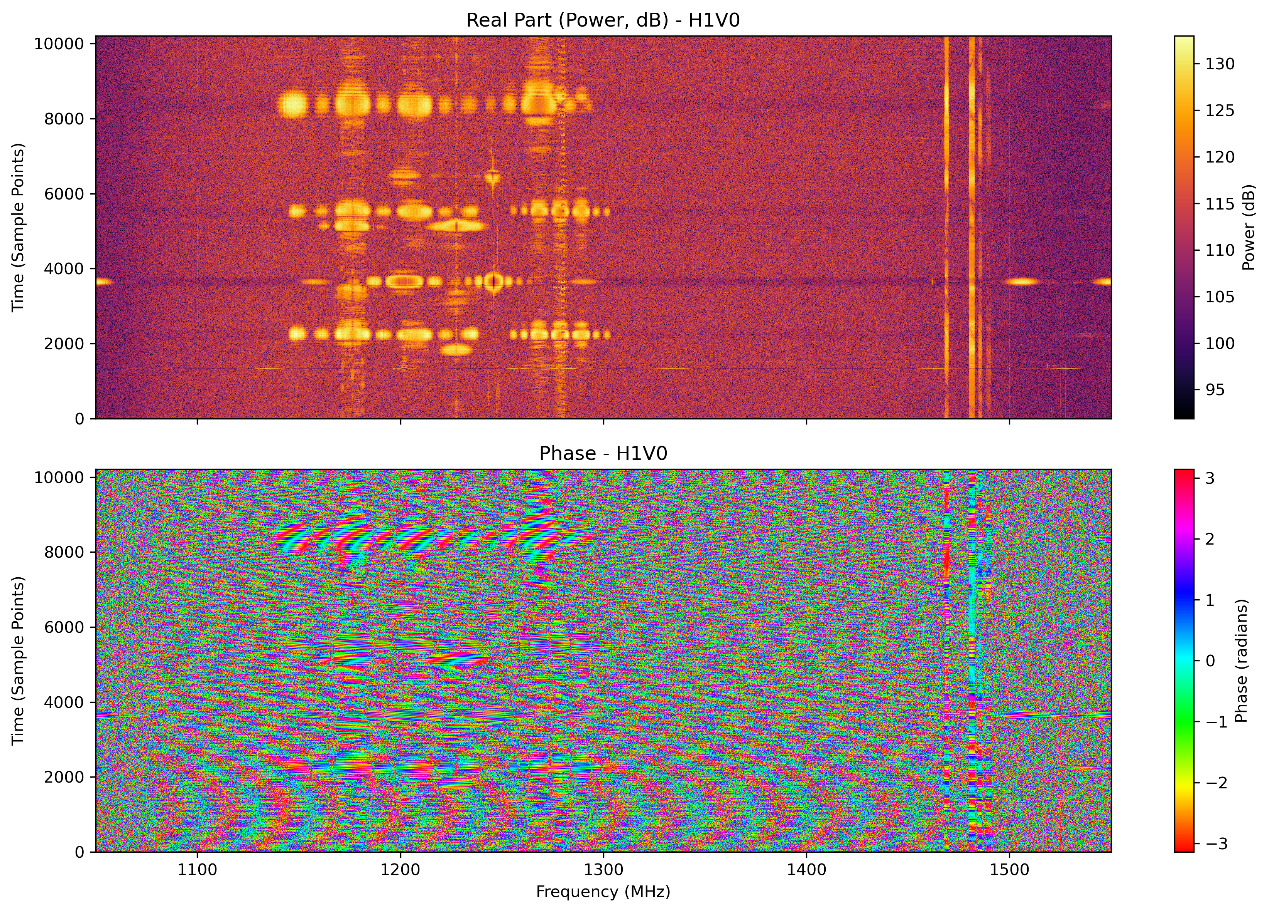}
\centering
\caption{Dynamic spectrum from a long integration on Cas~A for a single baseline. The upper panel shows power and the lower panel shows phase. Together they provide a compact diagnostic of both spectral occupancy and temporal behavior of the dominant RFI.
\label{fig:waterfall}}
\end{figure*}

This section is included as supporting context for the calibration and imaging results that follow. Because most QUEST antennas are located on a university campus, RFI assessment is a necessary commissioning step rather than an optional add-on. The correlator output can be used for that purpose directly. Figure~\ref{fig:waterfall} shows the visibility amplitude and phase for one representative baseline (H1V0) during an observation of Cassiopeia~A, a standard flux-density calibrator in radio astronomy \citep{Baars1977,Trotter2017}. The figure provides a practical view of the electromagnetic environment seen by the system before any detailed calibration is attempted.

The amplitude panel reveals strong contamination in the 1150--1300\,MHz and 1460--1500\,MHz ranges. In these bands the interfering signals exceed the astronomical target by tens of decibels and would dominate the visibilities if left untreated. The phase panel adds useful diagnostic information: approximately stationary phase patterns are consistent with fixed or slowly varying local emitters, whereas rapidly evolving curved fringes are more naturally associated with moving emitters such as aircraft- or satellite-related transmissions. By combining these signatures with national frequency-allocation information, we associate the main contaminated bands with the services listed in Table~\ref{tab:rfi_sources}.

\begin{deluxetable}{ll}
\tablecaption{RFI Frequency Band Allocation \label{tab:rfi_sources}}
\tablehead{\colhead{Frequency (MHz)} & \colhead{Interference Source}}
\startdata
1164-1300 & Satellite radio navigation \\
1164-1215 & Aeronautical radio navigation \\
1215-1300 & Satellite earth exploration \\
1467-1492 & Satellite broadcast \\
\enddata
\end{deluxetable}

The strongest contamination is associated with satellite-navigation signals in the 1164--1300\,MHz range. These features are both bright---typically $\sim 40$--60\,dB above the target source---and temporally structured, which makes them particularly challenging for downstream calibration. Applying the MAD-based flagger described in Section~\ref{sec:calibration} removes approximately $\sim 45\%$ of the samples in the contaminated bands overall, with the 1164--1300\,MHz region reaching a flagging fraction of about $\sim 65\%$.

For the purposes of this paper, the main result is that the correlator output is diagnostically useful: it separates the principal contaminated bands, indicates likely source classes, and provides the masks needed for subsequent calibration. In future work this information can be coupled to more targeted mitigation strategies, including notch filtering, adaptive cancellation, and wider baseline leverage from the remote station.

\section{Parameter Tuning and Performance Evaluation} \label{sec:performance}

This benchmark section is intended to identify a practical operating point for the current software and hardware stack rather than to claim a universal performance limit. Tests were carried out on the workstation summarized in Table~\ref{tab:config}.

\begin{deluxetable}{ll}
\tablecaption{Correlator Hardware and Software Configuration \label{tab:config}}
\tablehead{\colhead{Component} & \colhead{Specification}}
\startdata
CPU & Intel Xeon Silver 4210R \\
GPU & NVIDIA RTX 4090D (24GB) \\
RAM & 256GB (8 x 32GB DDR4) \\
Storage & 15TB NVMe SSD \\
Operating System & Ubuntu 22.04 \\
Core Libraries & Python 3.13 / CUDA 12.1 \\
\enddata
\end{deluxetable}

Two software parameters were scanned systematically: the processing batch size (\texttt{batch-packets}) and the I/O queue depth (\texttt{queue-max}). For each parameter combination we tracked four metrics: total throughput, GPU compute time, data-read time, and host-to-device (H2D) transfer time. The benchmark data set contained four channels, with each file containing 100 data packets and a size of approximately 2\,GB.

Figure~\ref{fig:performance} shows the resulting performance maps. On the tested platform the peak throughput reaches 1.51\,GB\,s$^{-1}$. In the four-antenna observing mode considered here, that throughput is sufficient for real-time operation: a 2\,s observation file is processed in about 1.2\,s. The best performance appears in a relatively narrow operating region, centered around \texttt{batch-packets} $=256$ and \texttt{queue-max} of about 12--20.

\begin{figure*}[ht!]
\includegraphics[width=0.8\textwidth]{performance.png}
\centering
\caption{Sensitivity of correlator performance to \texttt{batch-packets} and \texttt{queue-max}. The panels show overall throughput, GPU compute time, data-read time, and H2D transfer time. The best throughput on the present platform is obtained near \texttt{batch-packets}$=256$.
\label{fig:performance}}
\end{figure*}

The most useful interpretation of Figure~\ref{fig:performance} is that the pipeline is balanced, rather than purely compute-limited. Performance is set by the interaction of three stages---data read, H2D transfer, and GPU compute---and the optimum is reached when these stages overlap efficiently. The data-read metric also shows that I/O remains an important bottleneck in the current deployment, even though the servers are linked by IB, because the upper software layers rely on NFS and TCP for file access. When \texttt{batch-packets} is too small, the GPU is underutilized and the system becomes compute-bound; when the queue depth is poorly matched to the batch size, the overlap between I/O and compute degrades.

For the present platform we therefore adopt \texttt{batch-packets} = 256 and \texttt{queue-max} = 12 as the recommended operating point. This result is primarily a deployment guide for QUEST, but it also shows that the Python/CuPy implementation can sustain real-time small-array correlation without requiring a custom low-level code path.

\section{Visibility Data Processing and Calibration} \label{sec:calibration}

\begin{figure*}[ht!]
\includegraphics[width=0.8\textwidth]{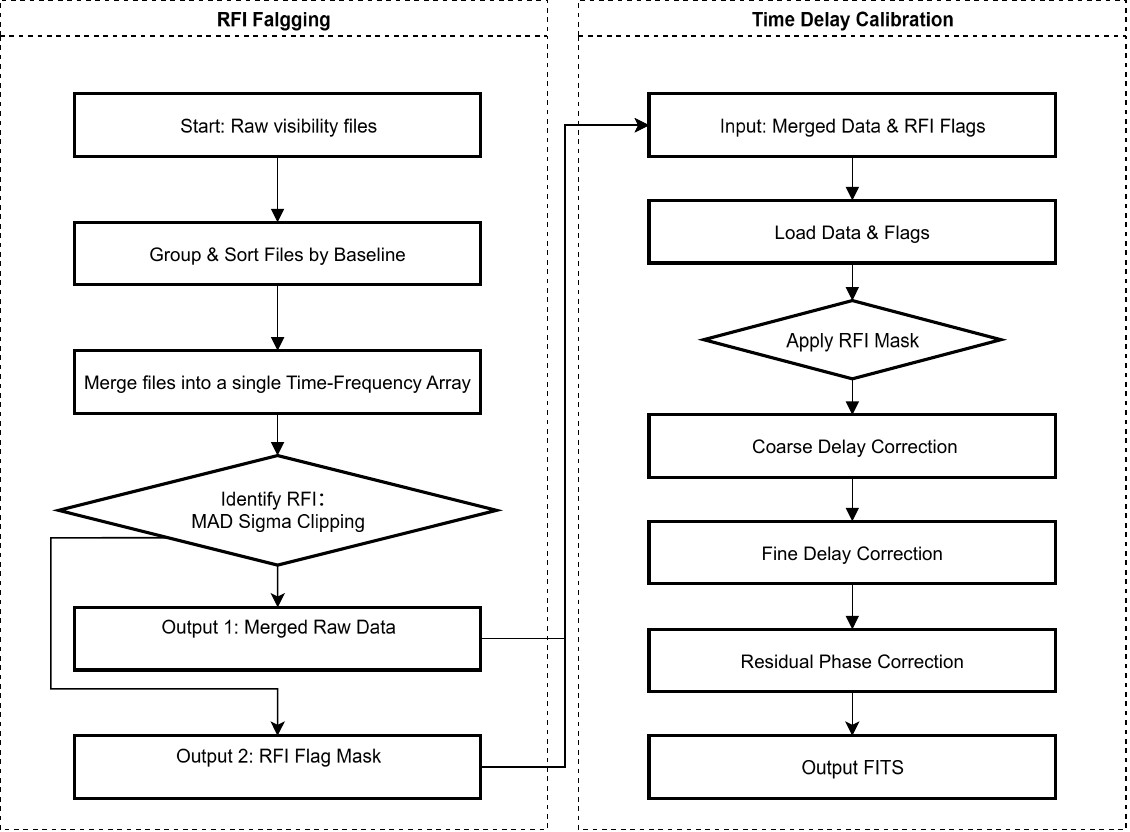}
\centering
\caption{Flowchart of the post-calibration steps. The first stage derives an RFI mask; the second stage applies the mask and performs successive delay and phase corrections.
\label{fig:cal_steps}}
\end{figure*}

The raw visibility output of the correlator is not yet suitable for imaging. Strong RFI, instrumental cable delays, and residual phase terms must first be removed. In that sense, the correlator is only one part of the full QUEST backend. The second part is a compact post-processing workflow that turns the correlator output into calibrated visibilities suitable for commissioning and initial synthesis imaging. Following the broader move toward software-defined calibration pipelines in radio astronomy \citep{Hunter_2023,ADEBAHR2022100514}, we implement a two-stage workflow consisting of RFI flagging followed by delay and phase calibration (Figure~\ref{fig:cal_steps}).

The first stage uses Median Absolute Deviation (MAD) sigma clipping to identify outliers in visibility amplitude. The threshold is
\begin{equation}
T = \text{median}(A) + \sigma \cdot \text{MAD}(A) \cdot 1.4826 ,
\end{equation}
where $A$ is the visibility amplitude, $\sigma$ is a user-defined threshold, and the factor 1.4826 converts MAD to the equivalent Gaussian standard deviation. In the results presented here we adopt $\sigma=5$, which provides conservative rejection of strong outliers while preserving most of the uncontaminated data. Over the range $\sigma\in[4,6]$, the behavior is qualitatively similar, with the exact flagging fraction depending on observing conditions. The resulting mask is written to a FITS file \citep{Wells1981} and then applied to the calibration stage.

Delay and phase calibration is performed in three serial steps:
\begin{enumerate}
    \item \textbf{Coarse group-delay correction.} The total delay $\tau_{\text{total}}$ is estimated from the peak of the delay spectrum, obtained by Fourier transforming the visibilities along the frequency axis.
    \item \textbf{Fine delay correction.} A weighted linear fit to the unwrapped phase--frequency relation in a clean band yields the residual slope $d\phi/d\nu$, from which
    \begin{equation}
    \tau_{\text{residual}} = \frac{1}{2\pi}\frac{d\phi}{d\nu}
    \end{equation}
    is derived.
    \item \textbf{Residual phase correction.} After the delay terms are removed, a frequency-independent but time-variable phase offset may remain. This term is estimated from the mean complex vector in the clean band and subtracted from all channels.
\end{enumerate}

\begin{figure}[h!]
\includegraphics[width=0.72\textwidth]{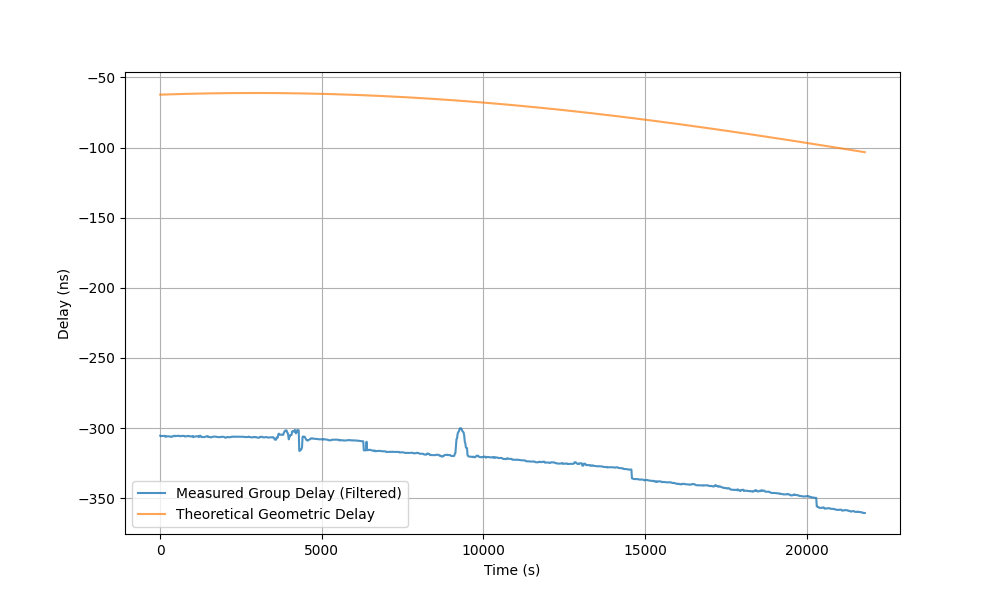}
\centering
\caption{Comparison of the measured group delay and the theoretical geometric delay. The offset between the two curves is dominated by non-geometric contributions that are removed by calibration.
\label{fig:delay_comp}}
\end{figure}

Figure~\ref{fig:delay_comp} compares the measured group delay after filtering with the theoretical geometric delay derived from the baseline vectors and source coordinates. The offset between the curves is dominated by instrumental plus atmospheric terms and is removed by the calibration procedure.

\begin{figure*}[ht!]
\includegraphics[width=0.5\textwidth]{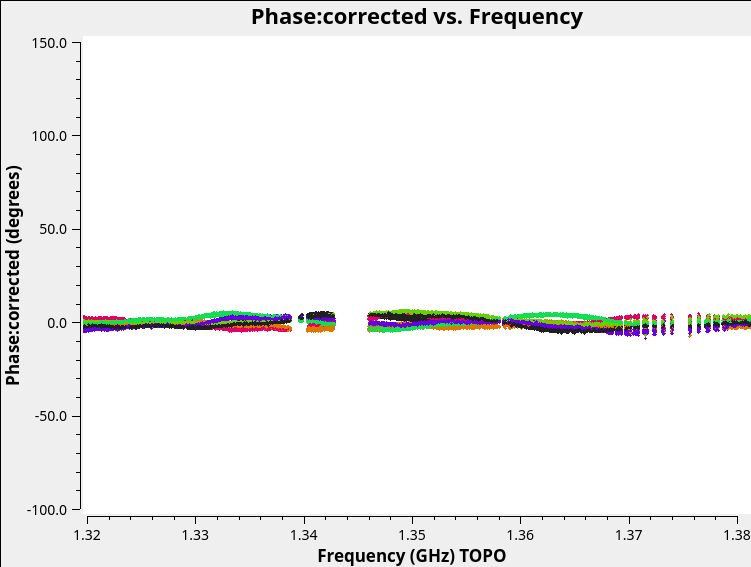}
\centering
\caption{Visibility phase as a function of frequency after application of the delay and residual-phase corrections. The dominant instrumental phase slope has been removed, leaving residual scatter of only a few degrees across the selected clean band.
\label{fig:cal_results}}
\end{figure*}

The effect of the full workflow is illustrated in Figure~\ref{fig:cal_results}. Prior to correction, the visibility phase is dominated by a large instrumental delay term and is therefore not usefully summarized by a single RMS value. After applying the coarse-delay, fine-delay, and residual-phase solutions, the phase-frequency relation becomes substantially flatter across the clean 1.32--1.38\,GHz band. The corrected phase scatter is only a few degrees overall (approximately $\sim 4^\circ$ RMS by visual estimate from the displayed band), which is sufficient for the imaging validation carried out in Section~\ref{sec:imaging}. A more complete closure-based error budget is deferred to future work.

\section{Imaging Validation} \label{sec:imaging}

The primary validation target of this paper is initial synthesis imaging from calibrated visibilities. We therefore use Cassiopeia~A as the main end-to-end test of the correlator, flagging, and calibration workflow. A shorter GNSS-based beam experiment is retained only as a supporting commissioning check, because it provides a first-look assessment of the field of view and reveals systematic offsets relevant to later hardware alignment.

\subsection{Synthesis Imaging of Cassiopeia A} \label{subsec:imaging}

The central validation step of this paper is synthesis imaging of Cassiopeia~A. This is not presented as a high-resolution science result: with only four antennas, the instantaneous $u$--$v$ coverage is sparse and the source is effectively unresolved. Instead, the aim is to test whether the correlator and calibration workflow can deliver visibilities that remain usable through gridding, Fourier inversion, and deconvolution. In that sense, the Cas~A result is the main evidence that the software is suitable for initial array commissioning.

\begin{figure*}[ht!]
\includegraphics[width=\textwidth]{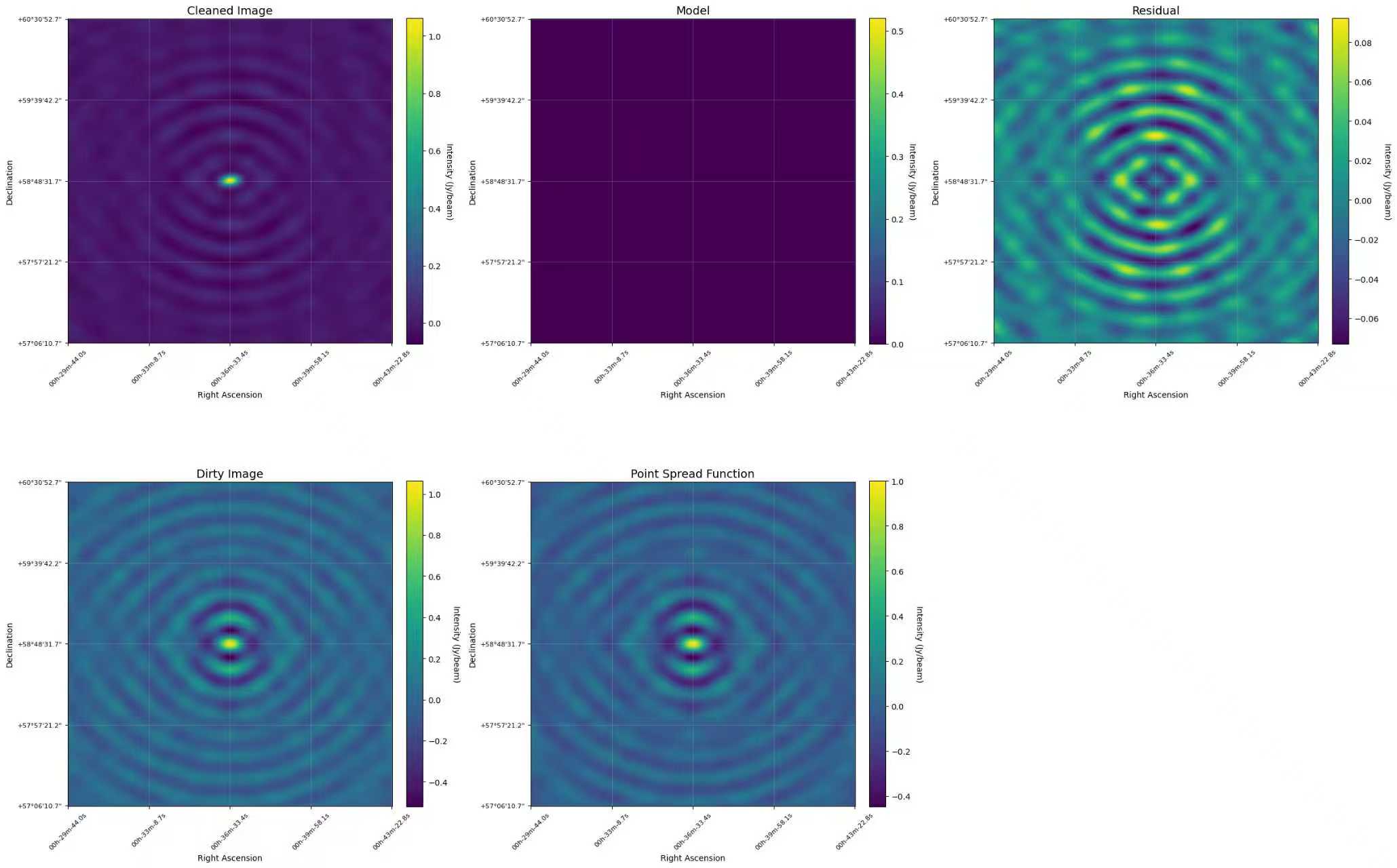}
\centering
\caption{Cassiopeia~A imaging with a four-antenna QUEST subarray. Top row: CLEANed image, CLEAN component model, and residual map. Bottom row: dirty image and synthesized beam (PSF). The source is recovered as a compact feature at the phase center, while the residual map retains ring-like structure associated with sparse $u$--$v$ coverage.
\label{fig:synthesis_sky}}
\end{figure*}

The calibrated visibilities from Section~\ref{sec:calibration} were gridded and transformed to produce the images shown in Figure~\ref{fig:synthesis_sky}. The figure displays the dirty map, the synthesized beam, the CLEANed image, the CLEAN component model, and the residual map. In this four-antenna configuration Cas~A is effectively unresolved, so the purpose of the figure is not to present a high-resolution science result but to demonstrate that the calibrated visibilities remain usable through the full imaging chain.

\begin{itemize}
    \item \textbf{Dirty map:} The inverse transform of the sampled visibilities shows Cas~A as the dominant peak at the phase center, but the surrounding field is strongly shaped by PSF sidelobes.
    \item \textbf{Point-spread function:} With only six instantaneous baselines, the synthesized beam has a compact main lobe and pronounced ring-like sidelobes, as expected for sparse $u$--$v$ coverage.
    \item \textbf{CLEANed map and residual:} Standard CLEAN deconvolution \citep{1974A&AS...15..417H} concentrates the source emission into a compact component at the phase center and lowers the image-domain background fluctuations, although structured residuals remain in the residual map.
\end{itemize}

From the plotted image scales, the peak brightness is close to $\sim 1$\,Jy\,beam$^{-1}$, the dirty-image background fluctuations are of order $\sim 0.1$\,Jy\,beam$^{-1}$, and the CLEANed-image background is reduced to a few $10^{-2}$\,Jy\,beam$^{-1}$. In other words, the image dynamic range improves from order $\sim 10$ in the dirty map to order $\sim 30$ after CLEAN, i.e., by a factor of a few. We therefore treat the Cas~A result as a commissioning-level imaging validation: it shows that the correlator plus calibration workflow produces visibilities that remain usable through gridding, inversion, and deconvolution, even though the present four-antenna configuration still leaves substantial PSF structure and does not yet support precision imaging metrics.

\subsection{Supporting GNSS Beam Check} \label{subsec:beam}

We treat the GNSS experiment as a supporting commissioning check rather than as a standalone beam-method paper. An estimate of the primary beam is still useful for field-of-view interpretation and later primary-beam correction, and GNSS signals in the L band provide a convenient moving probe for a rapid first-look measurement of the 4.5\,m QUEST antenna \citep{Berger_2026}. The purpose of this subsection is therefore limited: to determine whether the receiver plus analysis chain can recover a plausible beam envelope and reveal large systematic offsets that may affect subsequent observations.

\begin{figure*}[ht!]
\includegraphics[width=0.8\textwidth]{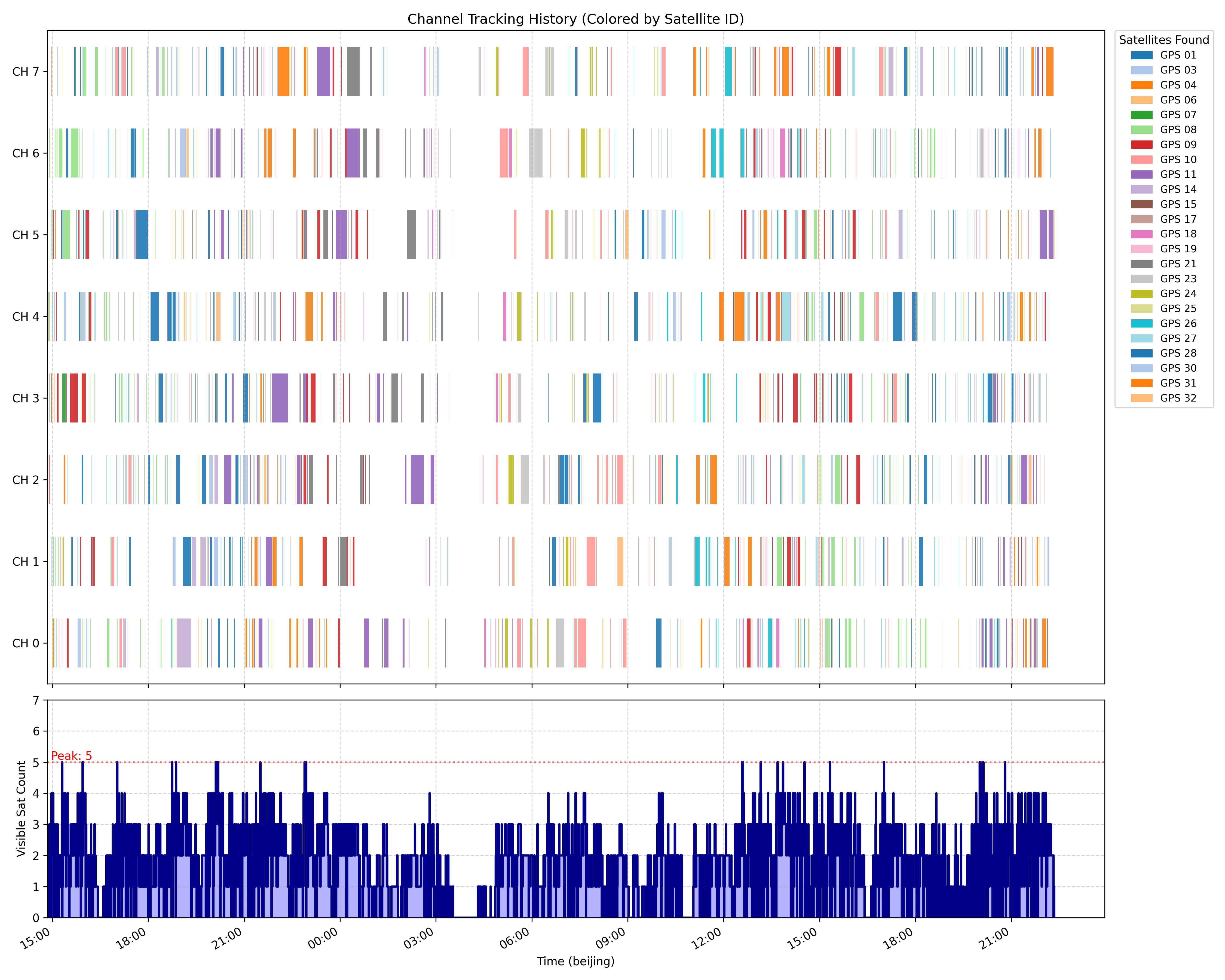} 
\centering
\caption{Tracking history of GPS L5 signals with a single zenith-pointing 4.5\,m antenna over 28\,hr. The upper panel shows which satellites were tracked by each channel; the lower panel gives the number of simultaneously visible satellites.
\label{fig:gnss_tracking}}
\end{figure*}

The antenna was fixed at the zenith ($90^\circ$ elevation), and a continuous 28\,hr observation of the GPS L5 band was recorded. Figure~\ref{fig:gnss_tracking} shows the channel-tracking history. Up to five satellites were tracked simultaneously, although the tracks are fragmented rather than continuous. Two effects likely dominate this behavior. First, the narrow main beam of a 4.5\,m dish means that many satellite passages occur through low-gain sidelobes rather than through the beam center, leading to a weak and rapidly varying carrier-to-noise ratio ($C/N_0$). Second, occasional data-buffer overflows in the high-throughput SDR chain can interrupt the continuous baseband stream and reset the receiver phase-locked loops.

Despite these interruptions, the aggregate data provide useful sampling over the local sky. Satellite coordinates were interpolated from SP3 precise ephemerides and transformed into local azimuth and zenith angle. Because the raw $C/N_0$ values include rapid fluctuations, deep fades, and multipath-induced nulls, direct averaging would bias the inferred beam downward. We therefore use an upper-envelope estimator: scatter points in the principal planes are binned in angle, and a high percentile (98th percentile in the present implementation) is used to trace the beam envelope.

\begin{figure*}[ht!]
\includegraphics[width=0.8\textwidth]{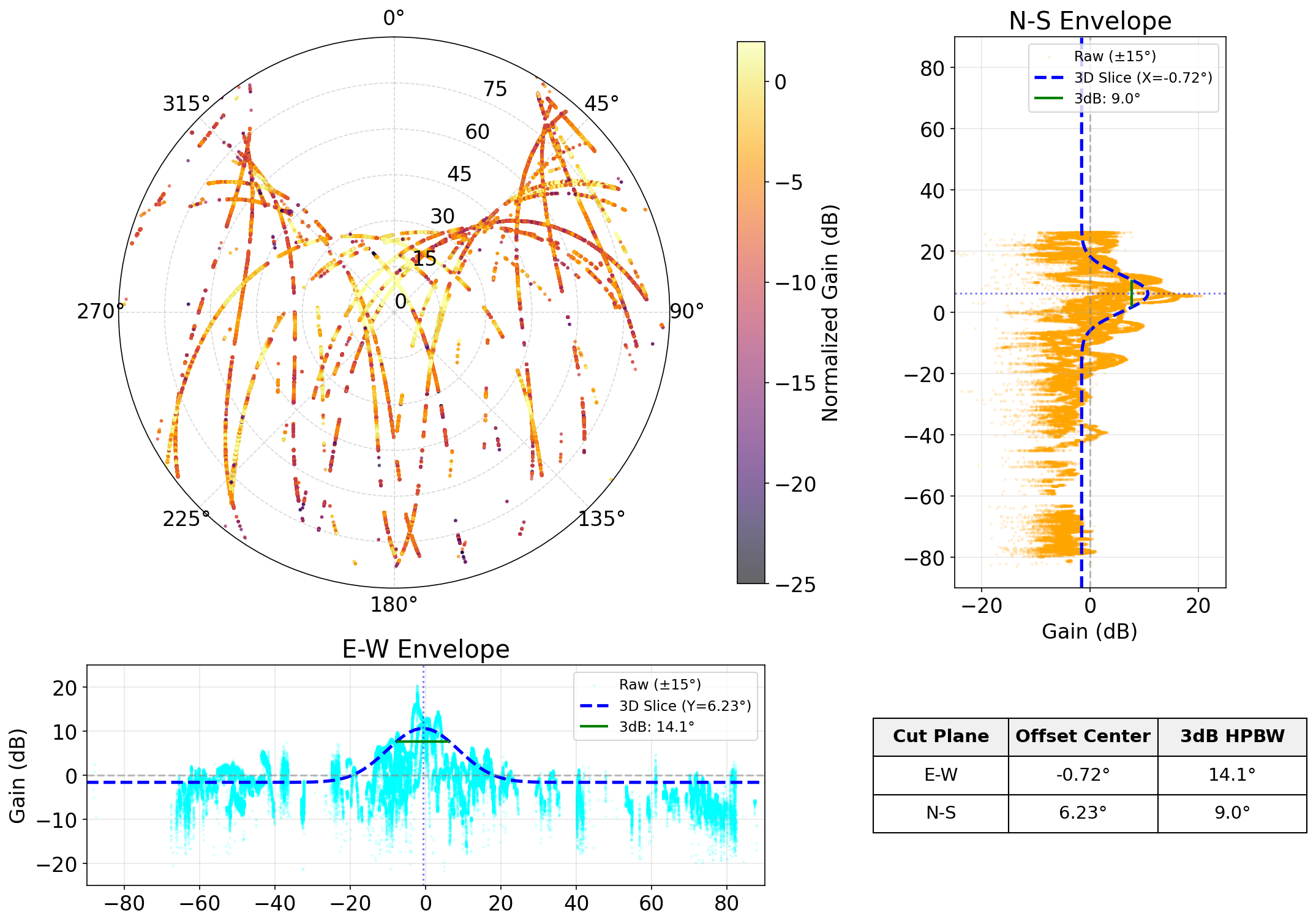}
\centering
\caption{Primary-beam estimate of a single 4.5\,m QUEST antenna derived from GNSS observations. The polar panel shows the reconstructed 2D beam pattern; the right and lower panels show the north--south and east--west cuts together with the percentile-based envelope estimate.
\label{fig:beam_map}}
\end{figure*}

Figure~\ref{fig:beam_map} shows that the recovered envelope is broadly consistent with the expected beam size, but not perfectly centered on the mechanical zenith. The east--west cut shows a small offset of about $-0.7^\circ$, whereas the north--south cut shows a larger offset of about $+6.2^\circ$ together with noticeable asymmetry. In the context of commissioning, this is a useful outcome rather than a failure: the method is sensitive enough to reveal mis-centering that may originate from pointing bias, feed displacement, timing offsets between the SDR clock and the ephemeris reference, or incomplete sampling of the exact beam peak.

Even with that offset, the recovered half-power beamwidth remains close to the expected $\sim 3.5^\circ$ value at 1.3\,GHz, which is adequate for the initial imaging exercise above. The main conclusion of this subsection is therefore that the GNSS-based procedure can provide a rapid first-look beam model while also highlighting systematic effects that should be addressed in later hardware alignment and timing calibration.

\section{Conclusion} \label{sec:conclusion}

We have presented a Python/CuPy software correlator for QUEST and evaluated it in the regime that motivated its design: small-array commissioning and initial imaging. The software emphasizes maintainability, configurability, and end-to-end diagnosis rather than the extreme throughput targets of very large production correlators. Within that scope, the implementation reaches a peak throughput of 1.51\,GB\,s$^{-1}$ on a single RTX~4090D and supports real-time operation for the four-antenna observing mode tested here.

The main result of the paper is that the correlator, together with a compact flagging and delay/phase-calibration workflow, produces visibilities that are adequate for initial synthesis imaging. This is illustrated with a four-antenna Cassiopeia~A test, in which the calibrated data remain usable through gridding and CLEAN deconvolution and show a reduction in image-domain background fluctuations from order $10^{-1}$ to a few $10^{-2}$\,Jy\,beam$^{-1}$. Supporting analyses of the local RFI environment and a GNSS-based first-look beam measurement show that the same software products are also useful for practical commissioning tasks.

Several limitations remain. The present validation is based on sparse four-antenna $u$--$v$ coverage; the beam check exposes a non-negligible north--south pointing offset; and some imaging and calibration metrics still merit a fuller uncertainty analysis. Future work should therefore prioritize quantitative residual-error statistics, larger-array deployments, and tighter integration of calibration into real-time operation. For the present paper, however, the conclusion is narrower and practical: the software is already sufficient to support small-array correlator commissioning and initial synthesis imaging on QUEST.

\begin{acknowledgments}
This work is supported by the National SKA Program of China (No. 2025SKA0110103, 2025SKA0140100), and the Natural Science Foundation of China (No. 12375108, 12275279, and 12333006). Jie Zhang is supported by the National Key R\&D Program of China (2023YFB4503305).
\end{acknowledgments}

\bibliography{sample701} 
\bibliographystyle{aasjournal}
\end{document}